\documentclass[a4paper,11pt]{article}
\usepackage{jcappub}
\usepackage{amsmath}
\usepackage{graphicx}
\usepackage{latexsym}
\usepackage{xspace}
\usepackage{color}

\makeatletter
\gdef\@fpheader{}
\makeatother

\newcommand{\ie}{i.e.\xspace}

\newcommand{\order}[1]{\mathcal{O}\!\left(#1\right)}
\newcommand{\Li}[1]{\mathrm{Li}_{#1}}

\DeclareMathOperator{\Ai}{Ai}
\DeclareMathOperator{\Bi}{Bi}

\newcommand{\dd}{\mathrm{d}}
\newcommand{\ee}{e}

\newcommand{\sss}[1]{{\scriptscriptstyle{#1}}}
\newcommand{\boldmathsymbol}[1]{{\ensuremath{\boldsymbol{#1}}}}

\newcommand{\uPl}{\mathrm{Pl}}

\newcommand{\uini}{\mathrm{ini}}

\newcommand{\us}{\mathrm{s}}
\newcommand{\uS}{\mathrm{S}}

\newcommand{\uT}{\mathrm{T}}

\newcommand{\usssS}{\sss{\uS}}

\newcommand{\usssT}{\sss{\uT}}

\newcommand{\usssPl}{\sss{\uPl}}

\newcommand{\nS}{n_\usssS}
\newcommand{\nT}{n_\usssT}
\newcommand{\alphaS}{\alpha_\usssS}
\newcommand{\alphaT}{\alpha_\usssT}

\newcommand{\uwkb}{\mathrm{WKB}}

\newcommand{\bmk}{\boldmathsymbol{k}}

\newcommand{\calH}{\mathcal{H}}
\newcommand{\calP}{\mathcal{P}}

\newcommand{\potS}{U_{_\uS}}
\newcommand{\potT}{U_{_\uT}}

\newcommand{\Mp}{M_\usssPl}

\newcommand{\Hstar}{H_*}

\newcommand{\astar}{a_*}

\newcommand{\cs}{c_\us}
\newcommand{\csini}{c_{\us \uini}}

\newcommand{\etastar}{\eta_*}
\newcommand{\nustar}{\nu_*}
\newcommand{\csstar}{c_{\us*}}
\newcommand{\epsonestar}{\epsilon_{1*}}
\newcommand{\epstwostar}{\epsilon_{2*}}
\newcommand{\epsthreestar}{\epsilon_{3*}}
\newcommand{\deltaonestar}{\delta_{1*}}
\newcommand{\deltatwostar}{\delta_{2*}}
\newcommand{\epsnstar}{\epsilon_{n*}}
\newcommand{\epsnponestar}{\epsilon_{n+1*}}

\newcommand{\wstar}{w_*}

\newcommand{\dia}{\sss{\diamond}}
\newcommand{\Hdia}{H_\dia}
\newcommand{\kdia}{k_\dia}
\newcommand{\etadia}{\eta_\dia}
\newcommand{\csdia}{c_{\us \dia}}
\newcommand{\epsonedia}{\epsilon_{1\dia}}
\newcommand{\epstwodia}{\epsilon_{2\dia}}
\newcommand{\epsthreedia}{\epsilon_{3\dia}}
\newcommand{\epsfourdia}{\epsilon_{4\dia}}
\newcommand{\epsfivedia}{\epsilon_{5\dia}}

\newcommand{\deltaonedia}{\delta_{1\dia}}
\newcommand{\deltatwodia}{\delta_{2\dia}}
\newcommand{\deltathreedia}{\delta_{3\dia}}
\newcommand{\deltafourdia}{\delta_{4\dia}}

\newcommand{\grav}{\star}
\newcommand{\Hgrav}{H_\grav}
\newcommand{\kgrav}{k_\grav}
\newcommand{\etagrav}{\eta_\grav}
\newcommand{\epsonegrav}{\epsilon_{1\grav}}
\newcommand{\epstwograv}{\epsilon_{2\grav}}

\newcommand{\aH}{\sss{\circledast}}
\newcommand{\kah}{k_\aH}
\newcommand{\etaah}{\eta_\aH}

\newcommand{\Dwkb}{D_{_\uwkb}}
\newcommand{\Awkb}{A_{_\uwkb}}

\newcommand{\Nstar}{N_*}

\subheader{}

\title{K-inflationary Power Spectra at Second Order}
 
\author[a]{J\'er\^ome Martin,}
\author[b]{Christophe Ringeval}
\author[a]{and Vincent Vennin}

\affiliation[a]{Institut d'Astrophysique de Paris, UMR
7095-CNRS, Universit\'e Pierre et Marie Curie, 98bis boulevard Arago,
75014 Paris (France)}

\affiliation[b]{Centre for Cosmology, Particle Physics and Phenomenology,
  Institute of Mathematics and Physics, Louvain University, 2 Chemin
  du Cyclotron, 1348 Louvain-la-Neuve (Belgium)}

\emailAdd{jmartin@iap.fr}
\emailAdd{christophe.ringeval@uclouvain.be}
\emailAdd{vennin@iap.fr}

\date{today}

\begin{document}

\abstract{Within the class of inflationary models, k-inflation
  represents the most general single field framework that can be
  associated with an effective quadratic action for the curvature
  perturbations and a varying speed of sound. The incoming flow of
  high-precision cosmological data, such as those from the Planck
  satellite and small scale Cosmic Microwave Background (CMB)
  experiments, calls for greater accuracy in the inflationary
  predictions. In this work, we calculate for the first time the
  next-to-next-to-leading order scalar and tensor primordial power
  spectra in k-inflation needed in order to obtain robust constraints
  on the inflationary theory. The method used is the uniform
  approximation together with a second order expansion in the Hubble
  and sound flow functions. Our result is checked in various limits in
  which it reduces to already known situations.}

\keywords{Cosmic Inflation, Slow-Roll, Cosmic Microwave Background}

\maketitle

\section{Introduction}

Inflation~\cite{Starobinsky:1980te, Guth:1980zm} (for reviews, see
Refs.~\cite{PPJPU,Martin:2003bt, Martin:2004um, Martin:2007bw,
  Sriramkumar:2009kg}), which is currently the leading paradigm to
describe the physical conditions that prevailed in the very early
Universe, is now entering a new phase. With the advent of new
high-accuracy cosmological data~\cite{Tonry:2003zg, Riess:2004nr,
  Riess:2006fw, AdelmanMcCarthy:2007aa, Abazajian:2008wr,
  Larson:2010gs, Komatsu:2010fb, Riess:2011yx, Hinshaw:2012fq,
  Bennett:2012fp, Hou:2012xq,
  Story:2012wx,Sievers:2013wk,Dunkley:2013vu}, among which are the
Planck data~\cite{2010AA520A9L}, one can hope to obtain very tight
constraints on the inflationary theory and even to pin-point the
correct model of inflation. In order to achieve this ambitious goal,
one must be able to compare the inflationary predictions to the
data. The problem is that the inflationary landscape is very
large~\cite{Barrow:1997fp} and that there is a whole zoo of different
models making different predictions. Moreover, for many of these
models, predictions can only be worked out by numerical methods. It is
therefore not obvious how to extract model-independent constraints on
the inflationary scenario.

How then should we proceed? Clearly, one can approach the problem step
by step and start with the simplest models. In other words, it seems
reasonable to consider more complicated models only if the data force
us to do so and tell us that the simplest models are not enough. Then
comes the question of identifying these models. One can convincingly
argue that slow-roll Single Field with a Minimal Kinetic term (SFMK)
scenarios are the simplest inflationary models since they are just
characterized by one function, the potential $V(\phi)$. In order to
establish their observational consequences, a possible approach is to
scan models one by one and calculate the predictions
exactly~\cite{Martin:2006rs, Bean:2007eh, Lorenz:2007ze,
  Martin:2010kz, Martin:2010hh, Easther:2011yq}, most of the time
numerically~\cite{Ringeval:2007am,Hazra:2012yn}\footnote{See for
  instance
  \url{http://theory.physics.unige.ch/~ringeval/fieldinf.html}.}. This
leads to an exact mapping of the inflationary landscape within this
class of scenarios but, given that the number of SFMK models remains
large, it would represent a huge effort. Another approach consists in
developing a scheme of approximation allowing us to derive analytical,
or semi-analytical, predictions. Although this is not always possible,
such a method is available for the SFMK models and one can explicitly
write a functional form for the primordial power spectrum of the
cosmological perturbations~\cite{Stewart:1993bc}, and even their
higher order correlation functions~\cite{Gangui:1993tt, Gangui:1994yr,
  Gangui:1999vg, Wang:1999vf, Maldacena:2002vr, Chen:2006nt}.

In fact, one can enlarge the class of what we consider as the simplest
models of inflation and assume that these ones are k-inflationary
scenarios. K-inflation~\cite{ArmendarizPicon:1999rj, Garriga:1999vw}
encompasses standard inflation and is more general since not only the
potential but also the kinetic term is now a free function. At the
perturbation level, the action for the comoving curvature perturbation
has a varying speed of sound and this describes all possible quadratic
terms within the effective field theory formalism~\cite{Cheung:2007st,
  Baumann:2011dt}. But, more interestingly, and despite the fact that
this class of scenarios is more complicated to analyze, a properly
generalized slow-roll approximation can still be used.

\subsection{State-of-the-art}

At this stage, it is interesting to recall the present status of the
techniques that enable us to calculate the two-point correlation
function for the primordial cosmological perturbations.

The spectrum of density perturbations during inflation was computed
for the first time in Refs.~\cite{Mukhanov:1985rz, Mukhanov:1988jd}
and for the gravity waves in Ref.~\cite{Starobinsky:1979ty}. Then, in
Ref.~\cite{Lucchin:1984yf}, it was realized that it can be evaluated
exactly in the case of power-law inflation. The first calculation at
first order in the so-called ``horizon flow parameters'' and using the
slow-roll approximation was performed in
Ref.~\cite{Stewart:1993bc}. This calculation was done for the SFMK
models. This is a fundamental result since it allows to connect the
deviations from scale invariance to the microphysics of
inflation. This result was re-derived using the Green function methods
in Ref.~\cite{Gong:2001he}, using the Wentzel-Kramers-Brillouin (WKB)
method in Ref.~\cite{Martin:2002vn} and using the uniform
approximation in Refs.~\cite{Habib:2002yi, Habib:2004kc}. In fact, the
Green function method of Ref.~\cite{Gong:2001he} made possible the
first determination of the scalar power spectrum at second order in
the ``horizon flow parameters''. Indeed, at second order, the mode
equation describing the evolution of the cosmological perturbations
can no longer be solved exactly, hence the need for a new method of
approximation. Higher order corrections were also obtained in
Ref.~\cite{Schwarz:2001vv}. The first derivation of the tensor power
spectrum at second order using the Green function method was presented
in Ref.~\cite{Leach:2002ar}. In Refs.~\cite{Casadio:2004ru,
  Casadio:2005xv}, it was also shown how to improve the WKB method by
adding more adiabatic terms. This improved WKB method has allowed a
re-derivation of the scalar and tensor power spectra at second order
and confirmed the results of the Green function approach.

After the advent of k-inflation, various attempts have been made to
derive the corresponding power spectra. The problem is complicated due
to the fact that density perturbations now propagate with a
time-dependent speed (the speed of sound). In Ref.~\cite{Wei:2004xx},
the Green function method has been used but with some
extra-assumptions on the behavior of the sound speed. The question
was also considered in Refs.~\cite{Shandera:2006ax,Bean:2007hc} but
the results obtained in those articles were not totally correct since
the sound speed was (implicitly) assumed to be constant which is not
the case in most of the k-inflationary scenarios (this result was also
used afterward in Ref.~\cite{Peiris:2007gz}). These works also missed
the influence of the sound speed in the tensor power spectrum due to
the shift between the scalar and tensor pivot
scales~\cite{Lorenz:2008et, Agarwal:2008ah}. The first fully
consistent result for the k-inflationary scalar power spectrum was
presented in Ref.~\cite{Kinney:2007ag}. The latter has been re-derived
using the uniform approximation in Ref.~\cite{Lorenz:2008et} together
with the first fully consistent calculation of the tensor power
spectrum at the same pivot scale. These spectra were compared to
Cosmic Microwave Background Anisotropy (CMB) data first in
Ref.~\cite{Lorenz:2008je}. However, all of these calculations have
been derived at first order only and no complete result at second
order exists in the literature.

The main purpose of this article is to close this gap and to derive
the slow-roll power spectra for the density and tensor perturbations
in k-inflation, at second order in the Hubble and sound flow
functions\footnote{Conforming to the modern usage, we will prefer the
  denomination of ``Hubble flow functions'' and ``sound flow
  functions'' to refer to the original, but confusing, appellation
  ``horizon flow parameters''. See Sec.~\ref{sec:kinflation} for the
  definition of the Hubble and sound flow functions.}. This
calculation is interesting for two reasons. Firstly, the second order
result is available for SFMK models and, for completeness, it should
also be done for the k-inflationary models. Secondly, according to the
"blue book"~\cite{bluebook}, Planck will measure the spectral index
with accuracy $\Delta \nS\simeq 0.005$. Even if one expects the Hubble
flow parameters to be less than $10^{-2}$, second order corrections
will be of order $10^{-4}$, that is to say relevant for high-accuracy
measurements of $\nS$ and/or estimation of the corresponding error
bars. Moreover, having at hand the second order terms allows to
marginalize over, a procedure that should always be carried on to get
robust Bayesian constraints on the first order terms.

Before moving to the calculation, let us briefly recall some
well-known results about k-inflation at the background and
perturbation levels.

\subsection{K-inflation in brief}
\label{sec:kinflation}

K-inflation corresponds to a class of models where gravity is
described by General Relativity and where the action for the inflaton
field is an arbitrary function, $P(\phi,X)$, the quantity $X$ being
defined by $X\equiv -(1/2)g^{\mu \nu}\partial_\mu \phi \partial_\nu
\phi$. This action can be written as
\begin{equation}
\label{eq:action}
S=\frac{\Mp^2}{2}\int \dd ^4 x \sqrt{-g}\left[
R+\frac{2}{\Mp^2}P\left(X,\phi\right)\right],
\end{equation}
where $\Mp$ is the reduced Planck mass. In fact, in order to satisfy
the requirements that the Hamiltonian is bounded from below and that
the equations of motion remain hyperbolic, the function $P(X,\phi)$
must satisfy the following two conditions~\cite{Bruneton:2006gf}
\begin{equation}
\label{eq:consistentconds}
\frac{\partial P}{\partial X}>0, \quad
2X\frac{\partial ^2P}{\partial X^2}+\frac{\partial P}{\partial X}>0.
\end{equation}
The general action~(\ref{eq:action}) includes standard inflation for
which $P=X-V(\phi)$, where $V(\phi)$ is the inflaton potential. This
class of model is in fact characterized by an arbitrary function of
$\phi$ only. K-inflation also includes the Dirac-Born-Infeld (DBI)
class of inflationary models~\cite{Alishahiha:2004eh}. For those, one
has $P=-T(\phi)\sqrt{1-2X/T(\phi)}+T(\phi)-V(\phi)$. This kind of
action typically appears in brane inflation and $T(\phi)$ is
interpreted as a warping function representing the bulk geometry in
which various branes can move. It is of course possible to find even
more complicated examples but, in the following, we will not need to
specify explicitly the function $P(X,\phi)$.

As in standard inflation, the dynamics of the background space-time
can be described by the Hubble flow functions $\epsilon_n$ defined by
\begin{equation}
\epsilon_{n+1}=\frac{\dd \ln \epsilon_n}{\dd N}\,, \qquad \epsilon_0
\equiv \dfrac{H_\uini}{H}\,,
\end{equation} 
where $N\equiv \ln (a/a_\uini)$ is the number of e-folds. Inflation
occurs if $\epsilon_1<1$ and the slow-roll approximation assumes that
all these parameters are small during inflation $\epsilon_n\ll 1$. Let
us notice that it is difficult to have an inflationary model without
such a condition because otherwise one would obtain a deviation from
scale invariance which would be too strong to be compatible with the
cosmological data (see however Ref.~\cite{Baumann:2011dt}).

At the perturbed level, we have density perturbations and gravity
waves. As usual, rotational perturbations are unimportant since they
quickly decay. Obviously, the tensorial sector of the theory is
standard since the gravitational part of~(\ref{eq:action}) is the
ordinary Einstein-Hilbert action. As a consequence, the equation of
motion for the amplitude $\mu_\bmk$ of gravity waves (re-scaled by a
factor $1/a$ for convenience, where $a$ is the
Friedman-Lemaitre-Robertson-Walker scale factor) takes the usual form,
namely
\begin{equation}
\label{eq:eommu}
\mu_\bmk''+\left[k^2-\potT(\eta)\right]\mu_\bmk=0,
\end{equation}
where $\eta $ is the conformal time and a prime denotes a derivative
with respect to $\eta$. The effective potential for the tensorial
modes can be written as $\potT=a^2H^2\left(2-\epsilon_1\right)$, \ie
only depends on the first Hubble flow function ($H=a'/a^2$ is the
Hubble parameter).

For the density perturbations, the situation is slightly more
complicated. One can show that the comoving curvature perturbation in
Fourier space, $\zeta_\bmk$, can be written in terms of a modified
Mukhanov-Sasaki variable $v_\bmk$ by means of the following
expression, $v_\bmk=(a\sqrt{\epsilon_1}) \zeta_\bmk/\cs$ (in Planck
units) where the quantity $\cs$ is defined by the following equation
\begin{equation}
\cs^2\equiv \frac{P_{,X}}{P_{,X}+2XP_{,XX}}\,,
\end{equation}
a subscript ``$,X$'' denoting differentiation with respect to
$X$. This quantity can be interpreted as the ``sound speed'' of
density fluctuations. Notice that, because of the two consistency
relations~(\ref{eq:consistentconds}), we have $\cs^2>0$. The fact that
$\cs$ is the sound speed can be most easily seen if one writes down
the equation of motion of the Mukhanov-Sasaki variable. It reads
\begin{equation}
\label{eq:eomv}
v_\bmk''+\left[\cs^2(\eta)k^2-\potS(\eta)\right]v_\bmk=0.
\end{equation}
This is similar to the equation of motion of a parametric
oscillator. The quantity $\potS$ is the effective potential for the
density perturbations and is a function of time only. As expected,
$\cs^2$ appears in front of the $k^2$ term, which is nothing but a
gradient term in Fourier space and this confirms its interpretation as
a time dependent sound speed. Since $\cs(\eta)$ is not known a priori,
one can introduce a second hierarchy of flow functions in order to
describe its behavior. Therefore, we define the sound flow functions
$\delta_n$'s by
\begin{equation}
\delta_{n+1}\equiv \frac{\dd \ln \delta _n}{\dd N}\,, \qquad \delta_0
\equiv \dfrac{\csini}{\cs}\,. 
\end{equation}
Consistent models of inflation are obtained if $\delta_n\ll 1$, that
is to say if the sound speed does not change too
abruptly~\cite{Kinney:2007ag, Lorenz:2008et}. A remark about
terminology is in order at this point. In terms of the Hubble and
sound flow functions, the effective potential for the density
perturbations can be expressed as
\begin{equation}
\potS=a^2H^2\left[2-\epsilon_1+\frac32 \epsilon_2
+\frac14 \epsilon_2^2-\frac12\epsilon_1\epsilon_2
+\frac12\epsilon_2\epsilon_3
+\left(3-\epsilon_1+\epsilon_2\right)\delta_1
+\delta_1^2+\delta_1\delta_2\right].
\end{equation}
The quantity $\potS$ depends on the $\epsilon_n$'s up to $\epsilon_3$
only and on the $\delta_n$'s up to $\delta_2$ only. Despite this last
property, it is important to remember that the above expression of
$\potS$ is \emph{exact} and that no approximation has been made at
this stage.

The cosmological observables we are interested in are the two point
correlation functions of the fluctuations, \ie in Fourier space, the
power spectra of both gravity waves and density perturbations:
\begin{equation}
\label{eq:powerspectra}
\calP_h = \frac{2k^3}{\pi^2}\left \vert \frac{\mu_\bmk}{a}\right\vert^2, 
\quad \calP_\zeta = \frac{k^3}{2\pi^2}
\left \vert \zeta_\bmk\right\vert^2=
\frac{k^3}{4\pi^2}
\frac{\cs^2\vert v_\bmk\vert^2}{\Mp^2a^2\epsilon_1}.
\end{equation} 
They have to be evaluated at the end of inflation and on large
scales. After the inflationary era, and for single field models, these
power spectra remain constant and can be directly used to compute
various observable quantities such as the CMB anisotropies or the
matter power spectrum. Our goal is now to integrate the equations of
motion of $\mu_\bmk$ and $v_\bmk$ in order to explicitly evaluate the
above power spectra.

This article is organized as follows. In the next section, after
having very quickly reviewed how the uniform approximation can be used
in the cosmological context, we apply it to the calculation of the
scalar and tensor primordial power spectra. Our results are discussed
Sec.~\ref{sec:conclusion}, in which we compare them, in the
appropriate limits, with the existing literature and we present our
conclusions.

\section{K-inflationary Power Spectra}
\label{sec:uniform}

\subsection{The Uniform Approximation}
\label{subsec:general}

In this section, we use the uniform approximation to calculate the
power spectrum of the density fluctuations in k-inflation, at second
order in the Hubble and sound flow functions. We have seen in the
previous section that the density perturbations in k-inflation
propagate with a time-dependent velocity $\cs(\eta)$. As the mode
equation can no longer be solved exactly in terms of Bessel functions
(even at first order for the sound flow functions), this prompts for the
use of new techniques. Here, we choose to work with the well-suited
uniform approximation~\cite{Habib:2004kc}. The idea is to rewrite the
effective potential according to $\potS=(\nu^2-1/4)/\eta^2$, an
equation which has to be understood as the definition of the function
$\nu(\eta)$. Then, we introduce two new functions
\begin{equation}
  g(\eta)\equiv
  \dfrac{\nu^2}{\eta^2} - \cs^2k^2, 
\qquad f(\eta)\equiv \frac{\vert \eta-\etastar \vert}{\eta-\etastar}
  \left \vert \frac{3}{2}\int _{\etastar}^{\eta}
    \dd \tau \sqrt{ g(\tau)}\right \vert^{2/3},
\end{equation}
where the turning point time $\etastar(k)$ is defined by the condition
$g(\etastar)=0$, that is to say $\etastar \equiv
-\nu(\etastar)/[k\cs(\etastar)]$. According to the uniform
approximation, the Mukhanov-Sasaki variable can then be expressed as
\begin{equation}
v_\bmk(\eta)=A_\bmk \left(\frac{f}{g}\right)^{1/4}\Ai\left(f\right)
+B_\bmk\left(\frac{f}{g}\right)^{1/4}\Bi\left(f\right),
\end{equation}
where the two constants $A_\bmk$ and $B_\bmk$ are fixed by the choice
of the initial conditions and where $\Ai$ and $\Bi$ denotes the Airy
function of the first and second kind respectively. Since one needs to
compute $v_\bmk$ on large scales, only the asymptotic behavior of
the Airy functions is needed and one arrives at a simpler formula,
namely
\begin{equation}
\lim_{\cs k \eta \rightarrow 0} v_\bmk(\eta) = \frac{B_\bmk}{g^{1/4}\pi^{1/2}}
\exp\left(\frac{2}{3}f^{3/2}\right).
\end{equation}
Here, the function $g(\eta)$ should be taken in its asymptotic limit,
\ie $g^{1/2} \simeq -\nu (\eta)/\eta $. Inserting the last equation
for $v_\bmk(\eta)$ into the formula~(\ref{eq:powerspectra}), one
obtains the following expression for $\calP_\zeta$
\begin{equation}
\label{eq:explicitPzeta}
\calP_\zeta=-\frac{k^3\left\vert B_\bmk\right\vert^2}{4\pi^3\Mp^2}
\frac{\eta\cs^2}{a^2\nu \epsilon_1}\ee ^{2\Psi},
\end{equation}
where we have defined $\Psi\equiv 2f^{3/2}/3$. One verifies that
$\calP_\zeta$ is positive definite since the conformal time is
negative during inflation. Therefore, the only thing which remains to
be done is to express the combination $\cs^2/(a^2\nu \epsilon_1)$ and
the quantity $\Psi$ at second order in the Hubble and sound flow
functions.

\subsection{Hubble and sound flow expansion}
\label{subsec:expansion}

The first step of the calculation consists in determining the
functions $a(\eta)$, $\cs(\eta)$, $\nu (\eta)$ and $\epsilon_1(\eta)$
at second order in the Hubble and sound flow functions. Here, we
first briefly explain the method in the case of the scale factor. By
definition, the conformal time is given by $\eta =-\int \dd t/a(t)$,
where $t$ is the cosmic time. By successive integrations by parts, one
can re-write $\eta $ as
\begin{equation}
\eta=\frac{1}{\calH}\left\{1+\epsilon_1+\epsilon_1^2+\epsilon_1\epsilon_2
-aH\int \frac{1}{a}\frac{\dd }{\dd a}
\left[\frac{1}{H}\left(\epsilon_1^2+\epsilon_1\epsilon_2\right)\right]\dd a
\right\},
\end{equation}
where $\calH\equiv a'/a$ is the conformal Hubble parameter. It is
important to stress that this equation is exact. In the last term, the
integrand is third order in the $\epsilon_i$. Indeed, differentiating
the term $1/H$ produces a $\epsilon_1$ which, multiplied with
$(\epsilon_1^2+\epsilon_1\epsilon_2)$, is third order. We also have to
differentiate expressions quadratic in the Hubble flow functions but,
since $\dd \epsilon_n/\dd N=\epsilon_n\epsilon_{n+1}$, this also gives
third order quantities. Therefore, the last term is
$\order{\epsilon^3}$ and can be dropped for a second order
calculation. In other words
\begin{equation}
\label{eq:calHsecond}
\calH =  -\frac{1}{\eta}\left(1+\epsilon_1+
\epsilon_1^2+\epsilon_1\epsilon_2\right) + \order{\epsilon^3}.
\end{equation}
In fact, this equation is not exactly what we want yet because,
although the second order terms $\epsilon_1^2$ and
$\epsilon_1\epsilon_2$ can be considered as constant in
time\footnote{Their derivative is indeed third order, \ie zero at the
  order at which we work.}, this is not the case for the term
$\epsilon_1$ which is first order. In order to render explicit the
time-dependence, let us notice that the equations defining the
Hubble-flow functions can also be written as $\dd \epsilon_n/\dd \eta
=\calH \epsilon_n\epsilon_{n+1}$. Given that
$\epsilon_n\epsilon_{n+1}$ is already a second-order term, we can just
replace $\calH$ with $-1/\eta$ in this expression and one gets
$\epsilon_n = \epsnstar-\epsnstar\epsnponestar\ln
\left(\eta/\etastar\right) + \order{\epsilon^3}$, where we have chosen
the integration constant such that this approximation is accurate
around the time $\etastar$ of the turning point. Inserting this
expression into Eq.~(\ref{eq:calHsecond}) gives
\begin{equation}
\calH=-\frac{1}{\eta}\left(1+\epsonestar+\epsonestar^2
+\epsonestar\epstwostar\right)
+\epsonestar\epstwostar\frac{1}{\eta}\ln
\left(\frac{\eta}{\etastar}\right) + \order{\epsilon^3},
\end{equation}
and, this time, the $\eta$-dependence of $\calH$ is explicit. This
equation can be further integrated leading to an expression for the
e-folds number $N$, namely
\begin{equation}
\label{eq:efold}
N-\Nstar=\ln \left(\frac{a}{\astar}\right)\simeq 
-\left(1+\epsonestar+\epsonestar^2+\epsonestar\epstwostar\right)
\ln \left(\frac{\eta}{\etastar}\right)
+\frac{1}{2}\epsonestar\epstwostar\ln ^2\left(\frac{\eta}{\etastar}\right).
\end{equation}
Finally, by exponentiation the above formula and by expressing the
constant $\astar\etastar$ in terms of $1/\Hstar$, one obtains the
following equation for the scale factor itself
\begin{align}
  a(\eta) & \simeq  -\frac{1}{\Hstar\eta}
  \biggl[1+\epsonestar+\epsonestar^2+\epsonestar\epstwostar
  -\left(\epsonestar+2\epsonestar^2+\epsonestar\epstwostar\right) \ln
  \left(\frac{\eta}{\etastar}\right) \nonumber \\ &
  +\frac{1}{2}\left(\epsonestar^2 +\epsonestar\epstwostar\right)\ln ^2
  \left(\frac{\eta}{\etastar}\right)\biggr].
\end{align}
We have reached our first goal, namely find an expression of $a(\eta)$
at second order in the Hubble flow parameters.

Let us now discuss how the expression of $\epsilon_1(\eta)$ can be
obtained. Let us notice that since $\epsilon_1$ is appearing in
Eq.~(\ref{eq:explicitPzeta}), we need to go to third order since a
term $1/\epsonestar$ will remain in front of the final expression of
$\calP_\zeta$. This can however be obtained using the above
formulas. Taylor expanding $\epsilon_1$ around $\Nstar$ one has
\begin{equation}
  \epsilon_1 = \epsonestar+\frac{\dd \epsilon_1}{\dd N}\biggl\vert_*\left(N-\Nstar\right)
  +\frac{1}{2}\frac{\dd ^2\epsilon_1}{\dd N^2}
  \biggl\vert_*\left(N-\Nstar\right)^2 + \cdots
\end{equation}
Using the fact that $\dd \epsilon_1/\dd N=\epsilon_1 \epsilon_2$, $\dd
^2\epsilon_1/\dd N^2=\epsilon_1\epsilon_2^2 +
\epsilon_1\epsilon_2\epsilon_3$ and the expression of the number of
e-folds at first order [see Eq.~(\ref{eq:efold}) above], one arrives
at
\begin{equation}
\epsilon_1=\epsonestar\left[1-\epstwostar\left(1+\epsonestar\right)
\ln \left(\frac{\eta}{\etastar}\right)+\frac12 \left(\epstwostar^2
+\epstwostar\epsthreestar\right)
\ln ^2\left(\frac{\eta}{\etastar}\right)\right]  +\order{\epsilon^4}.
\end{equation}
The very same method can be used to determine the second order
expression of the sound speed. Taylor expanding $\cs$ in e-fold gives
\begin{equation}
  \cs^2 = \csstar^2+\frac{\dd \cs^2}{\dd N} \biggl \vert_*\left(N-\Nstar\right)
  +\frac{1}{2}\frac{\dd ^2\cs^2}{\dd N^2} \biggl \vert_*
  \left(N-\Nstar\right)^2 + \cdots,
\end{equation}
and from the sound flow hierarchy one has $\dd \cs^2/\dd
N=-2\cs^2\delta_1$, $\dd ^2\cs^2/\dd
N^2=-2\cs^2\delta_1\delta_2+4\cs^2\delta_1^2$. Together with the
expression of $N-\Nstar$, it follows that
\begin{equation}
\label{eq:cssquaresecond}
\cs^2(\eta)=\csstar^2+2\csstar^2(\deltaonestar+\deltaonestar\epsonestar)
\ln \left(\frac{\eta}{\etastar}\right)
-\csstar^2\left(\deltaonestar\deltatwostar
-2\deltaonestar^2\right) 
\ln ^2\left(\frac{\eta}{\etastar}\right) +
\order{\epsilon^3,\delta^3}.
\end{equation}
As expected the coefficients of the logarithms are expressed 
in terms of the parameters $\delta_1 $ and $\delta_2$.

Finally, only the expression for $\nu(\eta)$ remains to be found. By
definition, one has $\nu^2=1/4+\eta^2 \potS (\eta)$, \ie Taylor
expanding everything from the previous formulas, one gets
\begin{equation}
\label{eq:nusecond}
\nu(\eta) =
\nustar-\left(\epsonestar\epstwostar+\frac{1}{2}\epstwostar\epsthreestar
  +\deltaonestar\deltatwostar\right)
\ln \left(\frac{\eta}{\etastar}\right) +
\order{\epsilon^3,\delta^3},
\end{equation}
with 
\begin{equation}
\nustar \equiv \frac{3}{2}+\epsonestar+\frac{1}{2}\epstwostar+\deltaonestar
+\epsonestar^2+\frac{11}{6}\epsonestar\epstwostar
+\frac{1}{6}\epstwostar\epsthreestar+\epsonestar\deltaonestar
+\frac{1}{3}\deltaonestar\deltatwostar.
\end{equation}

\subsection{Comoving Curvature Power Spectrum}
\label{subsec:calculationPzeta}

We have now determined explicitly the four functions appearing in the
expression of the power spectrum $\calP_\zeta$, see
Eq.~(\ref{eq:explicitPzeta}). It is straightforward, although lengthy,
to calculate, at second order, the relevant combination $\cs^2/(a^2\nu
\epsilon_1)$ appearing in that expression. Moreover, we must also find
$\Psi$. Upon using the expression of the function $g(\eta)$, one gets
\begin{equation}
\label{eq:defpsi}
\Psi =  \int_{\etastar}^{\eta}\dd \tau
\sqrt{\frac{\nu^2(\tau)}{\tau^2} - \cs^2(\tau)k^2}\,.
\end{equation}
Inserting Eqs.~(\ref{eq:cssquaresecond}) and (\ref{eq:nusecond}) into
the previous formula and expanding everything to second order, the
integrand in Eq.~(\ref{eq:defpsi}) reads
\begin{align}
\label{eq:integrandpsi}
\sqrt{\frac{\nu^2(\tau)}{\tau^2}-\cs^2(\tau)k^2}
&=-\frac{\nustar}{\tau}\left(1-\frac{\csstar^2k^2\tau^2}{\nustar^2}\right)^{1/2}
\nonumber \\ & +\frac{3}{2\nustar}\left(\epsonestar\epstwostar+\frac12
  \epstwostar \epsthreestar+\deltaonestar \deltatwostar\right)
\frac{1}{\tau}
\left(1-\frac{\csstar^2k^2\tau^2}{\nustar^2}\right)^{-1/2}
\ln \left(\frac{\tau}{\etastar}\right) \nonumber \\
& +\frac{\csstar^2}{\nustar}
\left(\deltaonestar+\epsonestar\deltaonestar\right) \frac{1}{\tau}
\left(1-\frac{\csstar^2k^2\tau^2}{\nustar^2}\right)^{-1/2} k^2\tau
^2\ln \left(\frac{\tau}{\etastar}\right)
\nonumber \\
&-\frac{\csstar^2}{2\nustar}
\left(\deltaonestar\deltatwostar-2\deltaonestar^2\right)
\frac{1}{\tau}
\left(1-\frac{\csstar^2k^2\tau^2}{\nustar^2}\right)^{-1/2} k^2\tau
^2\ln ^2\left(\frac{\tau}{\etastar}\right)
\nonumber \\
&+\frac{\csstar^4}{2\nustar^3}\left(\deltaonestar + \epsonestar
  \deltaonestar \right)^2 \frac{1}{\tau}
\left(1-\frac{\csstar^2k^2\tau^2}{\nustar^2}\right)^{-3/2} k^4 \tau^4
\ln^2 \left( \frac{\tau}{\etastar} \right).
\end{align}
Therefore, we have five different integrals to calculate in order to
evaluate the term $\Psi$. In the following, we write
\begin{equation}
\Psi=\sum_{i=1}^{i=5}I_i,
\end{equation}
and calculate each of the $I_i$ separately. Let us also notice that
the way Eq.~(\ref{eq:integrandpsi}) has been written is not yet fully
consistent since all the terms have to be expanded to
second-order. For instance, terms like
$(\deltaonestar+\epsonestar\deltaonestar)/\nustar$ (in front of the
second integral $I_2$) should clearly be expanded further on in order
to keep only second order expressions. For the moment, however, we
will be keeping them this way in order to maintain clarity. Only at
the end of the calculation these terms will be expanded.

Let us now calculate the five integrals. Defining $w\equiv \csstar
k\eta/\nustar$, which implies that $\wstar\equiv \csstar
k\etastar/\nustar=-1$, the first integral, $I_1$, can be calculated
exactly and reads
\begin{equation}
  I_1=-\nustar\left[\left(1-u^2\right)^{1/2}
    + \ln \vert u\vert 
    - \ln \left \vert 1+\left(1-u^2\right)^{1/2} \right \vert \right]
  \biggr \vert_{u=\wstar}^{u=w}.
\end{equation}
On large scales, $w$ approaches zero and one obtains
\begin{equation}
\lim _{w\rightarrow 0}I_1= -\nustar\left(1+\ln \vert w\vert -\ln 2\right).
\end{equation}
The second integral is slightly more complicated but can also be
carried out exactly. The result can be expressed as
\begin{align}
I_2&=\frac{3}{16\nustar}
\left(\epsonestar\epstwostar+\frac12 \epstwostar\epsthreestar
+\deltaonestar\deltatwostar\right)
\Biggl[4\ln ^2\vert u\vert 
-8\ln \vert u\vert \ln \left \vert 
\frac{1}{2}\left(1+\sqrt{1-u^2}\right)\right\vert 
\nonumber \\ &
+2\ln ^2\left \vert 
\frac{1}{2}\left(1+\sqrt{1-u^2}\right)\right\vert 
-4\, \Li2\left(\frac{1}{2}-\frac{\sqrt{1-u^2}}{2}\right)\Biggr]
\Biggr\vert_{u=\wstar}^{u=w},
\end{align}
where $\Li2$ denotes the Polygamma function of order two, or
dilogarithm function~\cite{Abramovitz:1970aa}. On large scales the
previous expression takes the form
\begin{equation}
\lim _{w\rightarrow 0}I_2=\frac{3}{16\nustar}
\left(\epsonestar\epstwostar+\frac12 \epstwostar\epsthreestar
+\deltaonestar\deltatwostar\right)
\left(4\ln ^2\vert w\vert -4\ln^2 2+\frac{\pi^2}{3}\right),
\end{equation}
where we have used $\Li2(0)=0$ and $\Li2(1/2)=\pi^2/12-(\ln ^22)/2$.
We notice that $I_1$ and $I_2$ are logarithmically divergent in the
limit $w\rightarrow 0$. We will see that this is not a problem and
that those terms cancel out in the final expression of
$\calP_\zeta$. This is expected since we know that the power spectrum
remains constant on larges scales, and as such an exact cancellation
of those terms constitutes a consistency check of the method. On the
contrary, the integrals $I_3$, $I_4$ and $I_5$ are convergent and can
be directly computed. They read
\begin{align}
I_3 &= \nustar \left(\deltaonestar+\epsonestar\deltaonestar\right)
\left(1-\ln 2\right), 
\quad I_4=-\frac{\nustar}{2}
\left(\deltaonestar\deltatwostar-2\deltaonestar^2\right)
\left(\frac{\pi^2}{12}-2+2\ln 2-\ln ^2 2\right), \nonumber \\ 
I_5&=\frac{\nustar}{2}
\left(\deltaonestar+\epsonestar\deltaonestar\right)^2
\left(2-\frac{\pi^2}{6}-2\ln 2+2\ln ^2 2\right).
\end{align}

This completes our calculation of the quantity $\Psi$ and we can now
evaluate the expression~(\ref{eq:explicitPzeta}). Collecting the
expressions of $a$, $\epsilon_1$, $\cs$ and $\nu$ established
previously, one gets $\cs^2/(a^2\nu \epsilon_1)$ that has to be
combined with $\ee^{2\Psi}$ using the above integrals. After some
lengthy but straightforward manipulations, one obtains
\begin{equation}
\begin{aligned}
\label{eq:pzetastar}
\calP_\zeta & = \frac{\Hstar^2 \left( 18 \ee^{-3}\right)}{8 \pi^2 \Mp^2
  \epsonestar \csstar} \Biggl[1+\left(-\frac{8}{3}+2\ln
  2\right)\epsonestar +\left(-\frac{1}{3}+\ln 2\right)\epstwostar
+\left(\frac{7}{3}-\ln 2\right)\deltaonestar \\ &
+\left(\frac{23}{18}-\frac{4}{3}\ln 2+\frac{1}{2}\ln ^2 2\right)
\deltaonestar^2 +\left(\frac{25}{9}-\frac{\pi^2}{24}-\frac{7}{3}\ln 2+
  \frac{1}{2}\ln ^2 2\right)\deltaonestar\deltatwostar \\ &
+\left(-\frac{25}{9}+\frac{13}{3}\ln 2-2\ln ^2 2\right)
\epsonestar\deltaonestar +\left(\frac{13}{9}-\frac{10}{3}\ln 2+2\ln ^2
  2\right)\epsonestar^2  \\ &
+\left(-\frac{2}{9}+\frac{5}{3}\ln 2-\ln ^2
  2\right)\epstwostar\deltaonestar
+\left(-\frac{25}{9}+\frac{\pi^2}{12}+\frac{1}{3}\ln 2+\ln^2 2\right)
\epsonestar \epstwostar  \\ &
+\left(-\frac{1}{18}-\frac{1}{3}\ln 2+\frac{1}{2}\ln^2 2\right)
\epstwostar^2 +\left(-\frac{1}{9}+\frac{\pi^2}{24}+\frac{1}{3}\ln 2
  -\frac{1}{2}\ln ^22\right)\epstwostar\epsthreestar\Biggr].
\end{aligned}
\end{equation}

Several remarks are in order at this stage. Firstly, in the above
calculation, we have assumed that the initial state of the
perturbations is the Bunch-Davies vacuum. This implies that $\vert
B_\bmk\vert ^2=\pi/2$. Notice that, in the context of k-inflation,
this is a non-trivial choice since, as discussed in
Ref.~\cite{Lorenz:2008et}, the time dependence of the sound speed
could be such that the adiabatic regime is not available
anymore\footnote{Let us notice however that one can still re-define a
  new time variable to absorb the $\cs$-dependence in the mode
  equation~\cite{Baumann:2011dt}. In terms of that new time variable,
  one could always set Bunch-Davies initial conditions for the scalar,
  but this would not be compatible with those of the tensor
  modes.}. In this paper, we assume that this does not occur and that
the function $\cs(\eta)$ is initially smooth enough. Secondly, as
announced above, all the time-dependent terms $\ln \vert w\vert$ have
canceled out and the expression of $\calP_\zeta$ is
time-independent. Thirdly, Eq.~(\ref{eq:pzetastar}) should be compared
with Eq.~(51) of Ref.~\cite{Lorenz:2008et}. These two expressions
coincide at first order, which is another indication that the above
formula for $\calP_\zeta$ is correct. Fourthly, in the overall
amplitude, we notice the presence of the factor $18\, \ee^{-3}$. As
explained in Refs.~\cite{Martin:2002vn} and \cite{Lorenz:2008et}, this
is typical in a approximation scheme based on the WKB method or its
extension (such as the uniform approximation). This leads to a $\simeq
10\%$ error in the estimation of the amplitude. In
Refs.~\cite{Casadio:2004ru, Casadio:2005xv}, it was shown that, by
taking into account higher order terms in the adiabatic expansion,
this shortcomings can easily be fixed. In that case, one obtains a new
overall coefficient which dramatically reduces the error in the
amplitude. As a consequence, we do not really need to worry about the
term $18\, \ee^{-3}$ and, for practical applications, one can simply
renormalize it to one.

Finally, the above expression of $\calP_\zeta$ depends on $\etastar$
which depends on $k$. Our goal is now to make this hidden scale
dependence explicit and to re-express the power spectrum at a unique
pivot scale defined by
\begin{equation}
  \kdia \etadia \equiv-\dfrac{1}{\csdia}\,.
\end{equation}
This is achieved by re-writing all the quantities appearing in the
power spectrum at a single time, $\eta =\etadia$. Technically, this
means that, say, $\epsonestar$ should be written as
$\epsonestar=\epsonedia-\epsonedia\epstwodia\ln
\left(\etadia/\etastar\right) +\order{\epsilon^3}$ and that the
dependence in $\etadia/\etastar$ should be replaced with a dependence
in $\kdia/k$. This is performed by making use of the relation between
the time $\etadia$ and $\etastar$:
\begin{equation}
\frac{\etastar}{\etadia}=\frac{\kdia}{k}\nustar\frac{\csdia}{\csstar}\,.
\end{equation}
Working out the previous equation at second order, one obtains that
\begin{align}
\ln \left(\frac{\etastar}{\etadia}\right)
&= \left(\ln \frac32 +\ln \frac{\kdia}{k}\right)
\left(1-\deltaonedia-\epsonedia\deltaonedia-\frac{2}{3}\epsonedia\epstwodia
-\frac{1}{3}\epstwodia\epsthreedia-\frac{2}{3}\deltaonedia\deltatwodia
+\deltaonedia^2\right)
\nonumber \\ & 
+\frac{2}{3}\epsonedia+\frac{1}{3}\epstwodia+\frac{2}{3}\deltaonedia
+\epsonedia\epstwodia-\frac{4}{9}\epsonedia\deltaonedia
+\frac{1}{9}\epstwodia\epsthreedia
+\frac{2}{9}\deltaonedia \deltatwodia
+\frac{4}{9}\epsonedia^2-\frac{1}{18}\epstwodia^2
\nonumber \\ &
-\frac{8}{9}\deltaonedia^2
-\frac{5}{9}\epstwodia\deltaonedia
+\frac12\deltaonedia \deltatwodia \ln^2 \frac{3}{2}
+\deltaonedia\deltatwodia \ln \frac32 \ln \frac{\kdia}{k}
+\frac{1}{2}\deltaonedia \deltatwodia \ln^2 \frac{\kdia}{k}.
\end{align}
This finally leads to one of the two main new results of this paper,
namely the expression of the scalar power spectrum in k-inflation at
second order in the Hubble and sound flow functions
\begin{align}
\label{eq:pzetafinal}
\calP_\zeta &=\frac{\Hdia^2\left(18\, \ee^{-3}\right)}
{8 \pi^2 \Mp^2\epsonedia \csdia}
\Biggl\{1-2(1+D)\epsonedia-D\epstwodia +(2+D)\deltaonedia
+\left(\frac{2}{9}+D+\frac{D^2}{2}\right)\deltaonedia^2
\nonumber \\ &
+\left(\frac{37}{18}+2D+\frac{D^2}{2}-\frac{\pi^2}{24}\right)\deltaonedia\deltatwodia
+\left(-\frac{8}{9}-3D-2D^2\right)\epsonedia \deltaonedia
+\left(\frac{17}{9}+2D+2D^2\right)\epsonedia^2
\nonumber \\ & 
+\left(\frac{5}{9}-D-D^2\right)\epstwodia\deltaonedia
+\left(-\frac{11}{9}-D+D^2+\frac{\pi^2}{12}\right)\epsonedia\epstwodia
+\left(\frac{2}{9}+\frac{D^2}{2}\right)\epstwodia^2
\nonumber \\ & 
+\left(\frac{\pi^2}{24}-\frac{1}{18}-\frac{D^2}{2}\right)
\epstwodia\epsthreedia
+\bigl[-2\epsonedia-\epstwodia+\deltaonedia
+(1+D)\deltaonedia^2
\nonumber \\ & 
+(2+D)\deltaonedia\deltatwodia
-(3+4D)\epsonedia\deltaonedia
+2(1+2D)\epsonedia^2-(1+2D)\epstwodia\deltaonedia
-(1-2D)\epsonedia\epstwodia
\nonumber \\ &
+D\epstwodia^2
-D\epstwodia\epsthreedia\bigr]\ln \frac{k}{\kdia}
+\biggl(2\epsonedia^2+\epsonedia\epstwodia+\frac{1}{2}\epstwodia^2
-\frac{1}{2}\epstwodia\epsthreedia
+\frac{1}{2}\deltaonedia^2+\frac{1}{2}\deltaonedia \deltatwodia
\nonumber \\ &
-2\epsonedia\deltaonedia
-\epstwodia\deltaonedia
\biggr)\ln ^2\frac{k}{\kdia}
\Biggr\},
\end{align}
where we have introduced the quantity $D$ defined by $D\equiv 1/3-\ln
3$. One easily checks that, at first order, this expression exactly
coincides with Eq.~(53) of Ref.~\cite{Lorenz:2008et}. More details in
the comparison of the above formula with the existing literature can
be found in Sec.~\ref{sec:conclusion}.

Using the method of Ref.~\cite{Schwarz:2004tz}, one can also deduce
the expression of the scalar spectral index which reads
\begin{align}
\nS-1& = -2\epsonedia-\epstwodia+\deltaonedia
-2\epsonedia^2-\left(2D+3\right)\epsonedia\epstwodia
+3\epsonedia\deltaonedia+\epstwodia\deltaonedia-D\epstwodia\epsthreedia
\nonumber \\ &
-\deltaonedia^2+\left(D+2\right)\deltaonedia\deltatwodia 
-2\epsonedia^3-\left(\frac{47}{9}+6D\right)\epsonedia^2\epstwodia
+5\epsonedia^2 \deltaonedia
\nonumber \\ &
+\left(-\frac{20}{9}-3D-D^2
+\frac{\pi^2}{12}\right)\epsonedia \epstwodia^2
+\left(-\frac{11}{9}-4D-D^2
+\frac{\pi^2}{12}\right)\epsonedia \epstwodia \epsthreedia
\nonumber \\ &
+\left(\frac{73}{9}+5D\right)\deltaonedia \epsonedia \epstwodia
-4 \epsonedia \deltaonedia^2
+\left(\frac{46}{9}+4D\right)\epsonedia \deltaonedia \deltatwodia
+\frac{4}{9}\epstwodia^2 \epsthreedia
\nonumber \\ &
+\left(-\frac{1}{18}-\frac{D^2}{2}+\frac{\pi^2}{24}\right)
\epstwodia \epsthreedia^2
+\left(\frac{5}{9}+2D\right)\deltaonedia \epstwodia \epsthreedia
\nonumber \\ &
+\left(-\frac{1}{18}-\frac{D^2}{2}+\frac{\pi^2}{24}\right)\epstwodia
\epsthreedia \epsfourdia
-\deltaonedia^2 \epstwodia
+\left(\frac{5}{9}+D\right)\deltaonedia \deltatwodia\epstwodia
+\deltaonedia^3
\nonumber \\ &
-\left(\frac{50}{9}+3D\right)\deltaonedia^2\deltatwodia
+\left(\frac{37}{18}+2D+\frac{D^2}{2}-\frac{\pi^2}{24}\right)
\deltaonedia\deltatwodia^2
\nonumber \\ &
+\left(\frac{37}{18}+2D+\frac{D^2}{2}-\frac{\pi^2}{24}\right)
\deltaonedia\deltatwodia\deltathreedia
\end{align}
At first order in the flow parameters, one recovers the standard
expression, \ie $\nS-1=-2\epsonedia-\epstwodia+\deltaonedia$. One can
also check that the second order corrections are similar to those
found in Ref.~\cite{Lorenz:2008et}. Here, for the first time, we have
given the formula of the spectral index at third order. This is of
course possible only because we have determined the overall amplitude
at second order. This also allows us to determine the higher order
corrections to the running and to the running of the running. For
instance, one can calculate $\alphaS$ at the fourth order and the
running of the running at the fifth order. Here, in order to
illustrate the efficiency of the method, we just present the
expression of $\alphaS$. It reads
\begin{align}
\alphaS &= -2\epsonedia\epstwodia-\epstwodia\epsthreedia
+\deltaonedia\deltatwodia
-6\epsonedia^2\epstwodia-\left(3+2D\right)\epsonedia\epstwodia^2
-\left(4+2D\right)\epsonedia\epstwodia\epsthreedia
+5\epsonedia\epstwodia\deltaonedia
\nonumber \\ & 
+4\epsonedia\deltaonedia\deltatwodia
-D\epstwodia\epsthreedia^2
-D\epstwodia\epsthreedia\epsfourdia
+2\deltaonedia\epstwodia\epsthreedia +\deltaonedia\deltatwodia\epstwodia
-3\deltaonedia^2\deltatwodia
+\left(2+D\right)\deltaonedia\deltatwodia^2
\nonumber \\ & 
+\left(2+D\right)\deltaonedia \deltatwodia\deltathreedia
-12\epsonedia^3\epstwodia
-\left(\frac{139}{9}+14D\right)\epsonedia^2\epstwodia^2
-\left(\frac{83}{9}+8D\right)\epsonedia^2\epstwodia\epsthreedia
\nonumber \\ & 
+21\deltaonedia\epsonedia^2\epstwodia
+9\deltaonedia\deltatwodia\epsonedia^2
+\left(-\frac{20}{9}-3D-D^2+\frac{\pi^2}{12}\right)\epsonedia\epstwodia^3
\nonumber \\ & 
+\left(-\frac{20}{3}-10D-3D^2+\frac{\pi^2}{4}\right)
\epsonedia\epstwodia^2\epstwodia
+\left(\frac{100}{9}+7D\right)\deltaonedia\epsonedia\epstwodia^2
\nonumber \\ & 
+\left(-\frac{11}{9}-5D-D^2+\frac{\pi^2}{12}\right)\epsonedia\epstwodia\epsthreedia^2
+\left(-\frac{11}{9}-5D-D^2+\frac{\pi^2}{12}\right)\epsonedia\epstwodia\epsthreedia
\epsfourdia
\nonumber \\ & 
+\left(\frac{127}{9}+7D\right)\deltaonedia\epsonedia\epstwodia\epsthreedia
-9\deltaonedia^2\epsonedia\epstwodia
+\left(\frac{137}{9}+9D\right)\deltaonedia\deltatwodia\epsonedia\epstwodia
-15\deltaonedia^2\deltatwodia\epsonedia
\nonumber \\ & 
+\left(\frac{64}{9}+5D\right)\epsonedia\deltaonedia\deltatwodia^2
+\left(\frac{64}{9}+5D\right)\epsonedia\deltaonedia\deltatwodia\deltathreedia
+\frac{8}{9}\epstwodia^2\epsthreedia^2
+\frac{4}{9}\epstwodia^2\epsthreedia\epsfourdia
\displaybreak[0] \nonumber \\ & 
+\left(-\frac{1}{18}-\frac{D^2}{2}+\frac{\pi^2}{24}\right)\epstwodia\epsthreedia^3
+\left(-\frac{1}{6}-\frac{3D^2}{2}+\frac{\pi^2}{8}\right)\epstwodia
\epsthreedia^2\epsfourdia
+\left(\frac{5}{9}+3D\right)\deltaonedia\epstwodia\epsthreedia^2
\displaybreak[0] \nonumber \\ & 
+\left(-\frac{1}{18}-\frac{D^2}{2}+\frac{\pi^2}{24}\right)
\epstwodia\epsthreedia\epsfourdia^2
+\left(-\frac{1}{18}-\frac{D^2}{2}+\frac{\pi^2}{24}\right)\epstwodia\epsthreedia
\epsfourdia\epsfivedia
\displaybreak[0] \nonumber \\ &
+\left(\frac{5}{9}+3D\right)\deltaonedia\epstwodia\epsthreedia\epsfourdia
-3\deltaonedia^2\epstwodia\epsthreedia
+\left(\frac{10}{9}+3D\right)\deltaonedia\deltatwodia\epstwodia\epsthreedia
-3\deltaonedia^2\deltatwodia\epstwodia
\nonumber \\ & 
+\left(\frac{5}{9}+D\right)\deltaonedia\deltatwodia^2\epstwodia
+\left(\frac{5}{9}+D\right)\deltaonedia\deltatwodia\deltathreedia\epstwodia
+6\deltaonedia^3\deltatwodia
-\left(\frac{118}{9}+7D\right)\deltaonedia^2\deltatwodia^2
\nonumber \\ & 
-\left(\frac{68}{9}+4D\right)\deltaonedia^2\deltatwodia\deltathreedia
+\left(\frac{37}{18}+2D+\frac{D^2}{2}-\frac{\pi^2}{24}\right)
\deltaonedia\deltatwodia^3
\nonumber \\ & 
+\left(\frac{37}{6}+6D+\frac{3D^2}{2}-\frac{\pi^2}{8}\right)
\deltaonedia\deltatwodia^2\deltathreedia
+\left(\frac{37}{18}+2D+\frac{D^2}{2}-\frac{\pi^2}{24}\right)
\deltaonedia\deltatwodia\deltathreedia^2
\nonumber \\ & 
+\left(\frac{37}{18}+2D+\frac{D^2}{2}-\frac{\pi^2}{24}\right)
\deltaonedia\deltatwodia\deltathreedia\deltafourdia
\end{align}
One can check that the second and third order corrections match the
expression already found in Ref.~\cite{Lorenz:2008et}. The fourth
order corrections represent a new result.

\subsection{Tensor Power Spectrum}
\label{subsec:calculationPh}

In this section, we repeat the previous analysis but for tensor
perturbations. Since the method is the same and, fortunately, the
calculations are easier, the details will be skipped. The main
difference between gravity waves and density perturbations is that
their effective potential is not the same, see Eqs.~(\ref{eq:eommu})
and~(\ref{eq:eomv}). This implies that the function $\nu(\eta)$ for
tensors is different from the one of the scalars. One gets for the
tensor
\begin{equation}
  \nu^2(\eta)=\frac{9}{4} + 3 \epsonestar + 4 \epsonestar^2  
  + 4 \epsonestar \epstwostar -3 \epsonestar \epstwostar
  \ln \left( \frac{\eta}{\etastar} \right) + \order{\epsilon^3}.
\end{equation}
As a consequence, the functions $g(\eta)$, $f(\eta)$, and hence
$\Psi$, are also different. Using the uniform
approximation to evaluate $\mu_\bmk$ and inserting the corresponding
formula into the expression of $\calP_h$ given by
Eq.~(\ref{eq:powerspectra}), one obtains
\begin{align}
\calP_h=& \frac{2\left(18\, \ee^{-3}\right)\Hstar^2}{\pi^2 \Mp^2}
\biggl[1+\left(-\frac{8}{3}+2\ln 2\right)\epsonestar
+\left(\frac{\pi^2}{12}-\frac{26}{9}+\frac{8}{3}\ln 2
-\ln^2 2\right)\epsonestar\epstwostar
\nonumber \\ & 
+\left(\frac{13}{9}-\frac{10}{3}\ln 2
+2\ln ^2 2\right)\epsonestar^2\biggr].
\end{align}
This equation is for the tensors what Eq.~(\ref{eq:pzetastar}) is for
the scalars. As explained before, one has still to make explicit the
scale dependence hidden in $\etastar$. In the case of tensors, the
pivot point is usually defined by $\kgrav \etagrav=-1$ since gravity
waves propagate at the speed of light. This leads to the following
expression for the power spectrum
\begin{align}
\calP_h=& \frac{2\left(18\, \ee^{-3}\right)\Hgrav^2}{\pi^2 \Mp^2}
\biggl\{1-2(1+D)\epsonegrav
+\left(\frac{17}{9}+2D+2D^2\right)\epsonegrav^2
+\biggl(-\frac{19}{9}+\frac{\pi^2}{12}
\nonumber \\ &
-2D
-D^2\biggr)
\epsonegrav \epstwograv
+\bigl[-2\epsonegrav+2(1+2D)\epsonegrav^2
-2(1+D)\epsonegrav \epstwograv \bigr]\ln \frac{\kgrav}{k}
\nonumber \\ &
+\left(2\epsonegrav^2-\epsonegrav \epstwograv\right)
\ln ^2\frac{\kgrav}{k}
\biggr\}.
\label{eq:phstdpiv}
\end{align}
Let us notice that, in order to obtain this relationship, we have used
the initial conditions for gravity waves $\vert
B_\bmk\vert=1/\Mp^2$. Otherwise, one notices the presence of the WKB
factor $18\, \ee^{3}$ and one can check that, at first order, it
coincides with the known expression for the tensor power spectrum. The
above formula, being expressed at the time $\etagrav$, is convenient
for SFMK models only, but not for k-inflation. Indeed, all parameters
here are functions evaluated at the time $\etagrav$ which is different
that the one at which the scalar power spectrum is calculated, namely
$\etadia$. It has become a common mistake to try fitting data with
both Eq.~\eqref{eq:pzetafinal} and Eq.\eqref{eq:phstdpiv} while
implicitly assuming that all Hubble and sound flow ``parameters'' are
the same. As we have explicitly shown before, they do differ and such
a fit would absolutely make no sense.

However, within slow-roll, one can re-express the tensor power
spectrum at the same pivot point as for the scalar power spectrum. As
before, each quantity in the tensor power spectrum should be
re-expressed at the scalar pivot point, as for instance $\epsonegrav =
\epsonedia -\epsonedia \epstwodia \ln \csdia +
\order{\epsilon^3,\delta^3}$. The quantity $\csdia $ appears because
it is present in the ratio of the tensor to scalar pivot points. It
follows that the final expression for the tensor power spectrum for
k-inflation is
\begin{align}
\label{eq:phfinal}
\calP_h=& \frac{2\left(18\ee^{-3}\right)\Hdia^2}{\pi^2 \Mp^2}
\biggl\{1-2(1+D-\ln \csdia)\epsonedia
+\biggl[\frac{17}{9}+2D+2D^2
+2\ln ^2 \csdia \nonumber \\ & - 2(1+2D)\ln \csdia
\biggr]\epsonedia^2
+\left[-\frac{19}{9}+\frac{\pi^2}{12}-2D-D^2
+2(1+D)\ln \csdia-\ln^2\csdia
\right]
\epsonedia\epstwodia
\nonumber \\ & 
+\biggl[-2\epsonedia+(2+4D-4\ln \csdia)\epsonedia^2+
(-2-2D+2\ln \csdia)\epsonedia\epstwodia\biggr]\ln \frac{\kdia}{k}
\nonumber \\ & 
+\left(2\epsonedia^2-\epsonedia\epstwodia\right)
\ln ^2\frac{\kdia}{k}
\biggr\},
\end{align}
where now ``diamonded'' terms are evaluated at the scalar pivot
point. This new formula is the second main result of the present
paper. It extends to second order the results of
Ref.~\cite{Lorenz:2008et}. As for the scalar modes, this expression
also allows us to calculate the tensor spectral index at third
order. One obtains
\begin{eqnarray}
\nT&=&-2\epsonedia-2\epsonedia^2+\left(-2-2D+2\ln \csdia\right)
\epsonedia \epstwodia -2\epsonedia^3
+\left(-\frac{38}{9}-6D+6\ln \csdia\right)\epsonedia^2\epstwodia
\nonumber \\ & &
+\left(-\frac{19}{9}-2D-D^2+\frac{\pi^2}{12}+2\ln\csdia
+2D\ln \csdia-\ln ^2 \csdia\right)\epsonedia\epstwodia^2
\nonumber \\ & &
+\left(-\frac{19}{9}-2D-D^2+\frac{\pi^2}{12}+2\ln\csdia
+2D\ln \csdia-\ln ^2 \csdia\right)\epsonedia\epstwodia\epsthreedia
\end{eqnarray}
Of course, at first order, one recovers the standard formula,
$\nT=-2\epsonedia$. We have already discussed before the relevance
of higher order corrections for Bayesian parameter estimation. Notice
that, in the case of primordial gravitational waves and as discussed
in Ref.~\cite{Kuroyanagi:2011iw}, another motivation is the
possibility of detecting them directly. Indeed, in that case, one
needs to estimate their power spectrum today and, due to the very
large lever arm between the cosmological scales and the smaller scales
where a direct detection can be performed, it is necessay to calculate
the power spectrum at the end of inflation very precisely. In this
context, higher order corrections become mandatory. Similarly, the
running of the tensors is obtained at fourth order and reads
\begin{align}
\alphaT & = -2 \epsonedia \epstwodia - 6 \epsonedia^2 \epstwodia + (-2
-2 D + 2 \ln \csdia) \epsonedia \epstwodia^2 + (-2 -2D + 2 \ln \csdia)
\epsonedia \epstwodia \epsthreedia
\nonumber \\
&  -12 \epsonedia^3 \epstwodia + \left(-\dfrac{112}{9} - 14D + 14 \ln\csdia \right)
\epsonedia^2 \epstwodia^2 + \left(-\dfrac{56}{9} - 8D + 8 \ln \csdia
\right) \epsonedia^2 \epstwodia \epsthreedia
\nonumber \\
& + \left[ -
  \dfrac{19}{9} -2D -D^2 + \dfrac{\pi^2}{12} + 2(1+D)\ln\csdia -
  \ln^2\csdia \right]\left(\epsonedia \epstwodia^3 + 3 \epsonedia
\epstwodia^2 \epsthreedia  \right. 
\nonumber \\ & \left. + \epsonedia \epstwodia
\epsthreedia^2 +\epsonedia \epstwodia \epsthreedia \epsfourdia\right).
\end{align}
Finally, one can also deduce the tensor to scalar ratio at the third
order. It reads
\begin{eqnarray}
r &=& 16\epsonedia\csdia\Biggl[1+2\epsonedia \ln \csdia +D\epstwodia
-\left(2+D\right)\deltaonedia
+\left(\frac{34}{9}+3D+\frac{D^2}{2}\right)\deltaonedia^2
\nonumber \\ & &
+\left(-\frac{37}{18}-2D-\frac{D^2}{2}+\frac{\pi^2}{24}\right)
\deltaonedia \deltatwodia
-\left(\frac{5}{9}+3D+D^2\right)\deltaonedia\epstwodia
+\left(-\frac{2}{9}+\frac{D^2}{2}\right)\epstwodia^2
\nonumber \\ & &
+\frac{1}{72}\left(4+36D^2-3\pi^2\right)\epstwodia\epsthreedia
+2\epsonedia^2\left(1+\ln \csdia  \right)\ln \csdia 
\nonumber \\ & & 
+\left(-\frac{28}{9}-3D-4\ln \csdia -2D\ln \csdia\right)\deltaonedia
\epsonedia
\nonumber \\ & &
+\left(-\frac{8}{9}+D+2\ln\csdia+4D\ln \csdia-\ln^2\csdia
\right)\epsonedia\epstwodia\Biggr].
\end{eqnarray}
As usual the leading term is proportional to $\epsonedia \csdia$ and
the above formula shows that the corresponding corrections depend on
the flow parameters but also on the sound speed.

\section{Discussion and Conclusions}
\label{sec:conclusion}

The power spectra of Eqs.~(\ref{eq:pzetafinal}) and~(\ref{eq:phfinal})
represent the main result of this article. There are the first
calculation, at second order in the Hubble and sound flow functions,
of the scalar and tensor power spectra in k-inflation within the
uniform approximation. In this section, we discuss our results and
check their consistency. In particular, in some limits, our
calculation should reproduce known results already derived in the
literature. As we show below, this is indeed the case.

We have seen before that the power spectrum is obtained as an
expansion around the pivot scale and that the most general expression
of $\calP_\zeta$ can be written as
\begin{equation}
\calP_\zeta(k)=\tilde{\calP}_\zeta(\kdia)\sum_{n\ge 0}\frac{a_n}{n!}
\ln ^n \frac{k}{\kdia},
\end{equation}
where $\tilde{\calP}_\zeta $ is the overall amplitude and the
coefficients $a_n$ are functions of the horizon flow parameters. The
expression of $a_n$ always starts at order $n$, \ie $a_0$ starts with
one, $a_1$ starts with a term of order $\order{\epsilon,\delta}$,
$a_2$ with a term of order
$\order{\epsilon^2,\delta^2,\epsilon\delta}$ and so on. As already
mentioned before, k-inflationary power spectra were determined at
first order in Ref.~\cite{Lorenz:2008et}. This means that the
expression found in that paper included only the first two terms,
proportional to $a_0$ and $a_1$. There is however a trick derived in
Ref.~\cite{Schwarz:2004tz} which allows us to determine some higher
order terms. Indeed, the power spectrum should not depend on the
choice of the pivot scale, which is arbitrary. As a consequence, one
can establish the following recursion relation
\begin{align}
a_{n+1}=\frac{\dd \ln \tilde{\calP}_\zeta}{\dd \ln \kdia}a_n
+\frac{\dd a_n}{\dd \ln \kdia}\,.
\end{align}
Given that $a_0$ was given at first order in
Ref.~\cite{Lorenz:2008et}, it was then possible to calculate $a_1$ up
to second order and $a_2$ to third order [see Eqs.~(64) and (65) in
that reference]. Therefore, one can compare those formulas to the
expression obtained in this article. One finds that they are the same,
indicating the consistency of our results.

Another way to verify the validity of our expressions is to take the
limit $\cs=1$ and to compare the resulting formulas to the results
already obtained in the literature for SFMK models. As mentioned in
the introduction, second order results were first obtained using the
Green function method in Ref.~\cite{Gong:2001he}. The corresponding
expression for the scalar power spectrum reads
\begin{align}
\label{eq:pzetagongstewart}
\calP_\zeta &=\frac{H^2} {8 \pi^2 \Mp^2\epsilon_1 }
\Biggl\{1-2(C+1)\epsilon_1-C\epsilon_2
+\left(2C^2+2C+\frac{\pi^2}{2}-5\right)\epsilon_1^2 \nonumber \\ &
+\left(C^2-C+\frac{7\pi^2}{12}-7\right)\epsilon_1\epsilon_2
+\left(\frac{C^2}{2}+\frac{\pi^2}{8}-1\right)\epsilon_2^2
+\left(-\frac{C^2}{2}+\frac{\pi^2}{24}\right) \epsilon_2\epsilon_3
\nonumber \\ & +\bigl[-2\epsilon_1-\epsilon_2 +2(2C+1)\epsilon_1^2
+(2C-1)\epsilon_1\epsilon_2 +C\epsilon_2^2
-C\epsilon_2\epsilon_3\bigr]\ln \frac{k}{\kah} \nonumber \\ &
+\biggl(2\epsilon_1^2+\epsilon_2\epsilon_2+\frac{1}{2}\epsilon_2^2
-\frac{1}{2}\epsilon_2\epsilon_3 \biggr)\ln ^2\frac{k}{\kah}
\Biggr\},
\end{align}
where the constant $C$ is defined by $C\equiv \gamma +\ln 2-2 \simeq
-0.7296$, $\gamma$ being the Euler constant, while the expression of
the gravity wave power spectrum can be written as
\begin{align}
\label{eq:pgwgongstewart}
\calP_h &=\frac{2 H^2}
{\pi^2 \Mp^2}
\Biggl\{1-2(C+1)\epsilon_1 
+\left(2C^2+2C+\frac{\pi^2}{2}-5\right)\epsilon_1^2
+\left(-C^2-2C+\frac{\pi^2}{12}-2\right)\epsilon_1\epsilon_2
\nonumber \\ &
+\bigl[-2\epsilon_1
+2(2C+1)\epsilon_1^2
-2(C+1)\epsilon_1\epsilon_2
\bigr]\ln \frac{k}{\kah}
+\biggl(2\epsilon_1^2-\epsilon_1\epsilon_2
\biggr)\ln ^2\frac{k}{\kah}
\Biggr\}.
\end{align}
In the two previous formulas~(\ref{eq:pzetagongstewart})
and~(\ref{eq:pgwgongstewart}), the Hubble flow functions are evaluated
at time $\etaah$ such that $a(\etaah)H(\etaah)=\kah$ which slightly
differs from the time $\kdia \etadia=-1$ (for $\csdia=1$) used in the
present paper. Therefore, if we want to compare
Eqs.~(\ref{eq:pzetafinal}) and~(\ref{eq:phfinal}) with $\csdia=1$ to
Eqs.~(\ref{eq:pzetagongstewart}) and~(\ref{eq:pgwgongstewart}), one
should first re-express the latter in terms of the Hubble flow
parameters evaluated at time $\kdia \etadia=-1$. In the following, in
order to simplify the discussion, we focus only on the scalar case but
the tensor case could be treated in the same manner. From the
definition of $\etaah$ one has $\etaah/\etadia = 1 + \epsonedia +
\epsonedia^2 + \epsonedia \epstwodia + \order{\epsilon^3}$. As consequence, in
Eqs.~(\ref{eq:pzetagongstewart}) and~(\ref{eq:pgwgongstewart}), one
should just replace $\epsilon_1$, $\epsilon_2$ with $\epsonedia$,
$\epstwodia$ and $H^2/\epsilon_1$ with
$\Hdia^2/\epsonedia(1+2\epsonedia^2 +\epsonedia\epstwodia)$. This
yields the following expression
\begin{align}
\label{eq:pzetamodifiedgs}
\calP_\zeta &=\frac{\Hdia^2}
{8 \pi^2 \Mp^2\epsonedia }
\Biggl\{1-2(C+1)\epsonedia-C\epstwodia 
+\left(2C^2+2C+\frac{\pi^2}{2}-3\right)\epsonedia^2
\nonumber \\ & 
+\left(C^2-C+\frac{7\pi^2}{12}-6\right)\epsonedia\epstwodia
+\left(\frac{C^2}{2}+\frac{\pi^2}{8}-1\right)\epstwodia^2
+\left(-\frac{C^2}{2}+\frac{\pi^2}{24}\right)
\epstwodia\epsthreedia
\nonumber \\ &
+\bigl[-2\epsonedia-\epstwodia
+2(2C+1)\epsonedia^2
+(2C-1)\epsonedia\epstwodia
+C\epstwodia^2
-C\epstwodia\epsthreedia\bigr]\ln \frac{k}{\kdia}
\nonumber \\ &
+\biggl(2\epsonedia^2+\epsonedia\epstwodia+\frac{1}{2}\epstwodia^2
-\frac{1}{2}\epstwodia\epsthreedia
\biggr)\ln ^2\frac{k}{\kdia}
\Biggr\},
\end{align}
that can be now compared to Eq.~(\ref{eq:pzetafinal}). As already
discussed, the overall amplitude differs by the WKB factor $18\,
\ee^{-3}$. We also notice that the terms in $D$ in
Eq.~(\ref{eq:pzetafinal}) exactly corresponds to the term in $C$ in
Eq.~(\ref{eq:pzetamodifiedgs}). For instance, the coefficient of
$\epsonedia^2$ in Eq.~(\ref{eq:pzetafinal}) contains a term $2D^2+2D$
while the coefficient of $\epsonedia^2$ in
Eq.~(\ref{eq:pzetamodifiedgs}) contains a $2C^2+2C$. One easily checks
that this is the rule for all first and second order terms. Provided
one substitutes $D$ with $C$, the first order term in the amplitude,
the coefficient of $\ln k/\kdia$ and the coefficient of
$\ln^2(k/\kdia)$ are identical. The only difference appears in the
second order terms in the amplitude. For instance, the coefficients of
$\epsonedia^2$ in Eq.~(\ref{eq:pzetafinal}) is $2D^2+2D+17/9$ while it
is $2C^2+2C+\pi^2/2-3$ in Eq.~(\ref{eq:pzetamodifiedgs}). But
$17/9\simeq 1.88$ and $\pi^2/2-3\simeq 1.93$ and, therefore, the two
terms are in fact numerically very close. The same is true for all the
other terms in the amplitude. Therefore, we conclude that our
result~(\ref{eq:pzetafinal}), specialized to SFMK models, is fully
consistent with Eq.~(\ref{eq:pzetamodifiedgs}) that comes from another
approximation scheme. This confirms its validity.

Let us now compare our result to the one of
Refs.~\cite{Casadio:2004ru, Casadio:2005xv} calculated with the help
of the WKB approximation. The scalar power spectrum obtained in those
articles reads
\begin{align}
\label{eq:pzetafabio}
\calP_\zeta &=\frac{H^2}
{8 \pi^2 \Mp^2\epsilon_1}\Awkb
\Biggl\{1-2(\Dwkb+1)\epsilon_1-\Dwkb\epsilon_2 
+\left(2\Dwkb^2+2\Dwkb-\frac{1}{9}\right)\epsilon_1^2
\nonumber \\ & 
+\left(\Dwkb^2-\Dwkb+\frac{\pi^2}{12}-\frac{20}{9}\right)
\epsilon_1\epsilon_2
+\left(\frac{\Dwkb^2}{2}+\frac{2}{9}\right)\epsilon_2^2
+\left(-\frac{\Dwkb^2}{2}+\frac{\pi^2}{24}-\frac{1}{18}\right)
\epsilon_2\epsilon_3
\nonumber \\ &
+\bigl[-2\epsilon_1-\epsilon_2
+2(2\Dwkb +1)\epsilon_1^2
+(2\Dwkb-1)\epsilon_1\epsilon_2
+\Dwkb\epsilon_2^2
-\Dwkb\epsilon_2\epsilon_3\bigr]\ln \frac{k}{\kah}
\nonumber \\ &
+\biggl(2\epsilon_1^2+\epsilon_1\epsilon_2
+\frac{1}{2}\epsilon_2^2
-\frac{1}{2}\epsilon_2\epsilon_3
\biggr)\ln ^2\frac{k}{\kah}
\Biggr\},
\end{align}
while the tensor power spectrum is given by the following formula
\begin{align}
\label{eq:pgwfabio}
\calP_h &=\frac{2 H^2}
{\pi^2 \Mp^2}\Awkb
\Biggl\{1-2(\Dwkb+1)\epsilon_1 
+\left(2\Dwkb^2+2\Dwkb-\frac{1}{9}\right)\epsilon_1^2
+\biggl(-\Dwkb^2-2\Dwkb
\nonumber \\ &
+\frac{\pi^2}{12}-\frac{19}{9}
\biggr)\epsilon_1\epsilon_2
+\bigl[-2\epsilon_1
+2(2\Dwkb+1)\epsilon_1^2
-2(\Dwkb+1)\epsilon_1\epsilon_2
\bigr]\ln \frac{k}{\kah}
\nonumber \\ &
+\left(2\epsilon_1^2-\epsilon_1\epsilon_2
\right)\ln ^2\frac{k}{\kah}
\Biggr\}.
\end{align}
In these equations, $\Awkb =18 \, \ee ^{-3}$ and $\Dwkb=1/3-\ln 3$,
that is to say exactly what was found by means of the uniform
approximation as $\Dwkb=D$.  As already mentioned,
Refs.~\cite{Casadio:2004ru,Casadio:2005xv} have shown that, by taking
the next order in the adiabatic approximation into account, one
obtains a new value for these two constants (in some sense, they are
renormalized), namely $\Awkb$ becomes $361/(18 \, \ee ^3)\simeq 0.99$
and $\Dwkb=7/19-\ln 3\simeq -0.7302$. In particular, the new value of
$\Dwkb$ is closer to the constant $C$ than the non-renormalized
one. Both Eqs.~(\ref{eq:pzetafabio}) and~(\ref{eq:pgwfabio}) are
evaluated at the pivot time $\etaah$ and have to be time-shifted to
$\etadia$ to be compared with our results. Proceeding as previously,
it is easy to show that this modifies the coefficients of
$\epsilon_1^2$ which now becomes $2\Dwkb^2+2\Dwkb+17/9$, and the
coefficient of $\epsilon_1\epsilon_2$ which becomes
$2\Dwkb^2-\Dwkb+\pi^2/12-11/9$. In other words
Eqs.~(\ref{eq:pzetafinal}) and Eq.~(\ref{eq:pzetafabio}) expressed at
$\etadia$ are exactly the same for $\csdia=1$. This is maybe not so
surprising considering the fact that the WKB and uniform
approximations are closely related methods.

A few words are in order about Ref.~\cite{Wei:2004xx}. Historically,
this is probably the first paper that attempted to evaluate the
k-inflationary power spectrum at second order in some equivalent of
the Hubble and sound flow functions used here. The method chosen is
the Green function expansion discussed before. However, a specific
form for the sound speed, which in the language of the present paper
would be a first order approximation of $\cs^2$, was also
postulated. Together with a $k-$dependence kept implicit, this makes
the comparison with the present work difficult. For this reason, we do
not investigate further this issue.

To conclude, let us briefly recap our main result and discuss
directions for future works. In this paper, using the uniform
approximation, we have calculated the scalar and tensor power spectra
in k-inflation, at second order in the Hubble and sound flow
parameters, see Eqs.~(\ref{eq:pzetafinal}) and~(\ref{eq:phfinal}). We
have carefully checked that, in the various limits where our
calculation reduces to known cases, consistent results are
obtained. The next step is clearly to use these power spectra in order
to constrain the values of the Hubble and sound flow parameters using
CMB observations. This was done in Ref.~\cite{Lorenz:2008je} but only
for the first order power spectra (since only this result was
available at that time). Given the on-going flux of high precision
data, such as those from the Planck satellite, the results obtained in
this article should be important to keep theoretical uncertainties at
a minimal level. In this way, as discussed in the introduction, one
may hope to obtain unprecedented information on the inflationary
scenario.

\bibliographystyle{JHEP}
\bibliography{kinf2nd}

\end{document}